\documentclass[aps,prl,amsmath,amssymb,reprint,superscriptaddress]{revtex4-2}

\usepackage{amsmath,charter,bm,graphicx,multirow}
\usepackage{graphicx}
\usepackage{dcolumn}
\usepackage{bm}
\usepackage{float}
\usepackage{color,soul}

\begin{document}

\title{Frustrated Ising charge correlations in the kagome metal ScV$_6$Sn$_6$}

\author{S.\,J.\,Gomez\,Alvarado}
\affiliation{Materials Department, University of California, Santa Barbara, California 93106, USA}

\author{G.\,Pokharel}
\affiliation{Materials Department, University of California, Santa Barbara, California 93106, USA}

\author{B.\,R.\,Ortiz}
\affiliation{Materials Science and Technology Division, Oak Ridge National Laboratory Oak Ridge, Tennessee 37831, USA}

\author{Joseph\,A.\,M.\,Paddison}
\affiliation{Neutron Scattering Division, Oak Ridge National Laboratory, Oak Ridge, Tennessee 37831, USA}

\author{Suchismita\,Sarker}
\affiliation{Cornell High Energy Synchrotron Source, Cornell University, Ithaca, New York 14853, USA}

\author{J.\,P.\,C.\,Ruff}
\affiliation{Cornell High Energy Synchrotron Source, Cornell University, Ithaca, New York 14853, USA}

\author{Stephen D. Wilson}
\affiliation{Materials Department, University of California, Santa Barbara, California 93106, USA}

\begin{abstract}

\noindent Here we resolve the real-space nature of the high-temperature, short-range charge correlations in the kagome metal ScV$_6$Sn$_6$. 
Diffuse scattering appears along a frustrated wave vector $\textbf{q}_H=(\frac{1}{3},\frac{1}{3},\frac{1}{2})$ at temperatures far exceeding the charge order transition $T_{CO}=92$~K, preempting long-range charge order with wave vectors along $\textbf{q}_{\bar{K}}=(\frac{1}{3},\frac{1}{3},\frac{1}{3})$.  Using a combination of real space and reciprocal space analysis, we resolve the nature of the interactions between the primary out-of-plane Sc-Sn chain instability and the secondary strain-mediated distortion of the in-plane V kagome network. A minimal model of the diffuse scattering data reveals a high-temperature, short-ranged ``zig-zag'' phase of in-plane correlations that maps to a frustrated triangular lattice Ising model with antiferromagnetic interactions and provides a real-space understanding of the origin of frustrated charge order in this material.  
\end{abstract}

\maketitle

\noindent 

One of the motivators in the search for new kagome-based compounds is to develop materials whose Fermi surfaces have electron fillings close to either hopping interference-driven flat bands or saddle points endemic to the kagome band structure.  For the latter features, nesting between the Van Hove singularities created by the saddle points is predicted to stabilize a number of unconventional states \cite{PhysRevLett.110.126405,PhysRevB.87.115135}, ranging from bond order to orbital magnetism to superconductivity \cite{PhysRevB.104.045122, PhysRevB.85.144402, Zhou2022}.  While one class of compounds of the form $A$V$_3$Sb$_5$ ($A$ = K, Rb, Cs) shows evidence for these instabilities \cite{Wilson2024}, finding other classes of kagome-based metals with a similar fillings and band structures is highly desirable.

The $R$V$_6$Sn$_6$ ($R$ = rare earth) family of V-based kagome metals are one such class with the potential of hosting correlated behavior due to the presence of Van Hove singularities near the Fermi level \cite{PhysRevB.104.235139, PhysRevLett.127.266401}. However, unlike the related $A$V$_3$Sb$_5$ family of V-based kagome metals, none of the $R$V$_6$Sn$_6$ members show evidence of superconductivity \cite{doi:10.7566/JPSJ.90.124704, PhysRevB.104.235139, PhysRevMaterials.6.083401, PhysRevMaterials.6.105001}, and only one member has shown evidence of a charge order instability \cite{PhysRevLett.129.216402}. This unique member is ScV$_6$Sn$_6$, and the origins of charge order in this compound are an area of active study---namely, investigating whether the charge order stems from a similar mechanism as that in $A$V$_3$Sb$_5$ compounds or whether its mechanism is distinct. 

The long-range distortion in the charge-ordered state of ScV$_6$Sn$_6$ below $T_{CO}=92$~K forms along a single $\textbf{q}_{\bar{K}}=(\frac{1}{3}, \frac{1}{3}, \frac{1}{3}$) \cite{PhysRevLett.129.216402}, a wave vector distinct from the three-\textbf{q} order in $A$V$_3$Sb$_5$ compounds \cite{Wilson2024}.  Rather than the dominant in-plane motion of V atoms in the kagome net of $A$V$_3$Sb$_5$ compound, the largest atomic displacements in ScV$_6$Sn$_6$ derive from motion of the Sc and Sn atoms that displace orthogonal to the kagome planes; however, as these atoms displace, they also induce a smaller amplitude motion of V atoms within the kagome planes. The coupling of these two instabilities can lead to a frustration of long-range charge order, which is suggested by the observation of short-range charge correlations that persist far above $T_{CO}$ \cite{pokharel_frustrated_2023, Korshunov2023}.

An oddity in the short-range charge fluctuations in ScV$_6$Sn$_6$ is that they appear along a wave vector $\textbf{q}_H=(\frac{1}{3}, \frac{1}{3}, \frac{1}{2}$) that differs from $\textbf{q}_{\bar{K}}$ of the long-range order state.  At finite frequencies, phonons also curiously soften along this same $\textbf{q}_H$ \cite{Cao2023,Korshunov2023} prior to the first order onset of $T_{CO}$.  Density functional theory (DFT) calculations predict several, nearly degenerate, types of lattice reconstructions in ScV$_6$Sn$_6$ \cite{tan_abundant_2023, PhysRevMaterials.8.014006}, and, due to this near degeneracy between lattice states, proposals of an order-disorder type phase transition have been put forward \cite{PhysRevB.109.L121103,PhysRevMaterials.8.014006}.  Phonon calculations further show an imaginary mode that is nearly flat at $\textbf{q}_H$ \cite{hu2023kagome}, and strong anharmonic phonon-phonon interactions are proposed as a mechanism of stabilizing higher temperature charge order fluctuations along $\textbf{q}_H$ \cite{wang2024origin}. 

One possible microscopic, real-space picture of the short-range charge correlations at $\textbf{q}_H$ is the correlated motion of Sn-Sc-Sn ``trimers" along the $c$ axis (Fig. 1) \cite{pokharel_frustrated_2023}. The smaller Sc ion on the $R$ site of this compound promotes this trimerization via the motion of the Sn atoms \cite{Meier2023}, and a Peierls-like instability forms along the Sn-Sc-Sn-Sn chain where each trimer has the degree of freedom to either push Sn atoms into or away from the kagome plane, providing an Ising-like degree of freedom.  Assuming an overall energy penalty for two neighboring trimers displacing with the same phase into (or away) from a single kagome plane within a bilayer, frustrated correlations are expected along the $\textbf{q}_H=(\frac{1}{3}, \frac{1}{3}, \frac{1}{2}$).  Experimentally resolving whether these high-temperature, short-range correlations indeed arise from the frustrated Ising-like motion of these trimers however remains an open challenge.

\begin{figure*}[t]
    \includegraphics[width=1.0\linewidth]{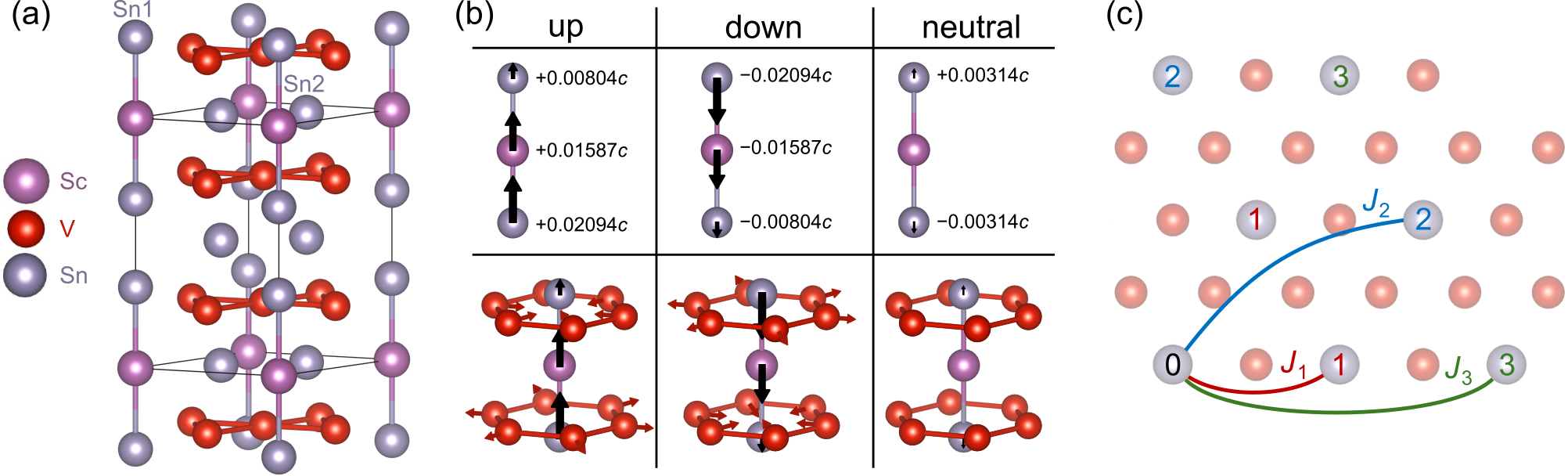}
    \caption{(a) The room-temperature structure of ScV$_6$Sn$_6$, showing Sc-Sn bonds and V-V bonds. (b) Depiction of the three trimer displacement modes, their amplitudes as observed in the low-temperature $\sqrt{3}\times\sqrt{3}\times3$ charge-ordered structure, and the expected corresponding displacements of the local vanadium sites. (c) A schematic of the first three nearest-neighbor pathways for trimer-trimer interactions.}
    \label{fig:structure}
\end{figure*}

In this paper, we perform an in-depth analysis of the diffuse x-ray scattering at the wave vectors $\textbf{q}_H$ where short-range charge order appears in ScV$_6$Sn$_6$.  Analysis reveals that the diffuse scattering originates from the correlated motion of Sn-Sc-Sn trimer blocks, similar to those that form below $T_{CO}$ in the long-range ordered state.  We further demonstrate that the in-plane diffuse scattering present above $T_{CO}$ can be captured via a minimal model of two-dimensional Ising-like displacements of Sn atoms that are frustrated via repulsive strain fields across the V-based kagome lattice.  Our combined three-dimensional difference pair distribution function (3D-$\Delta$PDF) and Monte Carlo analysis demonstrates that the high temperature pseudogap features reported in ScV$_6$Sn$_6$ derive from conduction electrons coupling to a frustrated charge instability that maps to a network of Ising spins on a two-dimensional triangular lattice antiferromagnet.

 \begin{figure}[b]
     \includegraphics[width=1.0\linewidth]{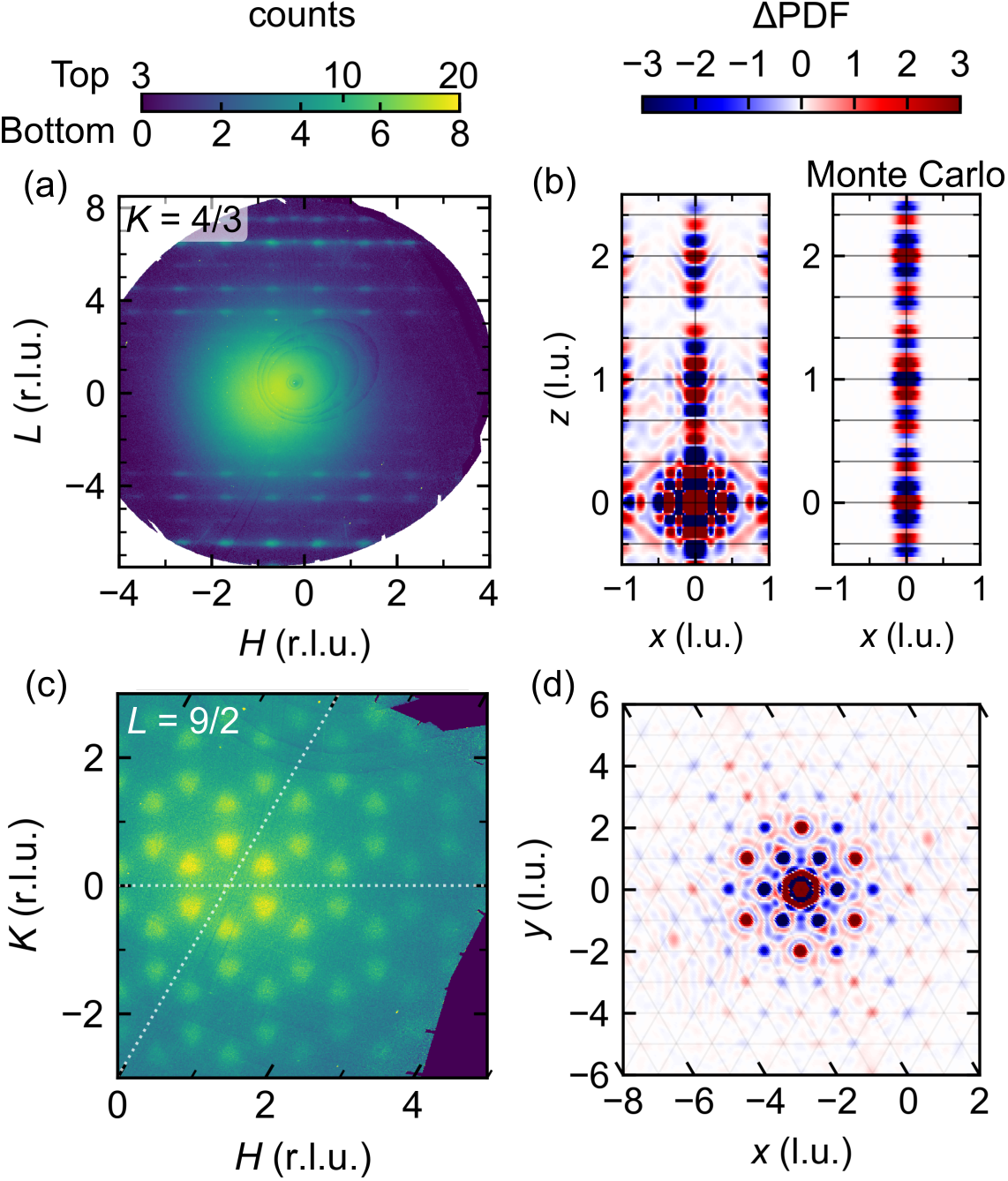}
     \caption{(a) X-ray data collected in the $(H, \frac{4}{3}, L)$ scattering plane. (b) Experimentally generated and simulated 3D-$\Delta$PDF in the $xz$ plane. (c) X-ray data collected in the $(H, K, \frac{9}{2})$ scattering plane. (d) Experimental 2D-$\Delta$PDF in the $xy$ plane. The experimental data were collected at $T=125$~K.}
     \label{fig:xyxzPDFcombined}
 \end{figure}
 
To explore the nature of the high-temperature charge correlations in ScV$_6$Sn$_6$, crystals were grown using previously reported flux growth techniques \cite{Meier2023, pokharel_frustrated_2023}. Synchrotron x-ray diffraction experiments were performed at the QM$^2$ beamline at the Cornell High Energy Synchrotron and data analyzed via the \textsc{NeXpy}/\textsc{NXRefine} and \textsc{nxs-analysis-tools} software suites \cite{nexpy, GomezAlvarado2024_nxs} and Monte Carlo analysis of the diffuse scattering was performed using the \textsc{DISCUS} software suite \cite{proffen_discus_1999}. Estimated interaction strengths were obtained from fits to the data performed using the software package \textsc{Spinteract} \cite{Paddison_2023}. Further details are provided in the Supplemental Material accompanying this paper \cite{SuppInfo}. 

 \begin{figure*}[]
    \includegraphics[width=1.0\linewidth]{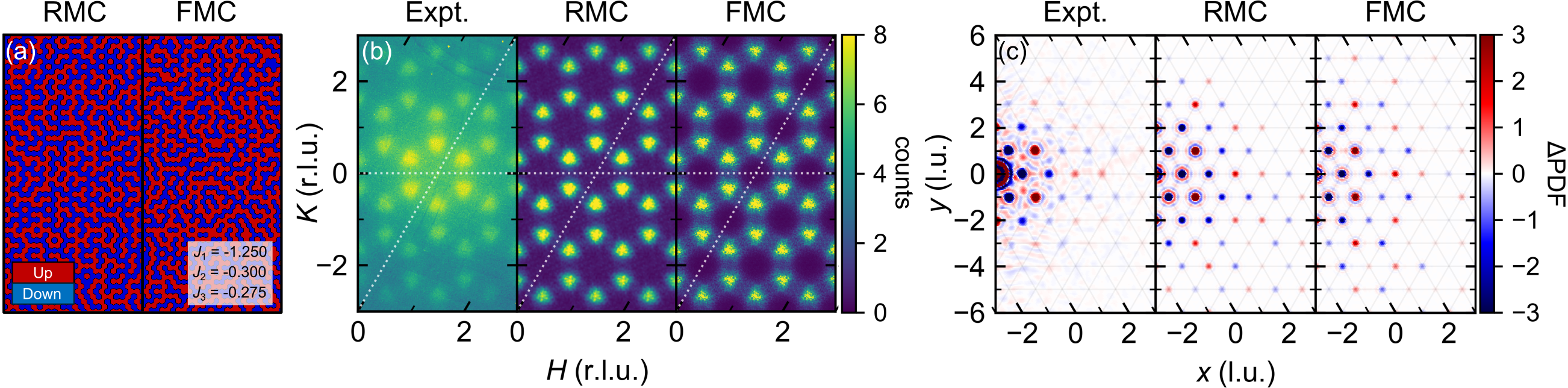}
    \caption{
    (a) Real-space $xy$ plane obtained from reverse and forward Monte Carlo modeling (only Sn1 sites shown). The two displacement types (up/down) are highlighted in red and blue. 
    (b) Comparison of the diffuse scattering in the $(H,K,\frac{9}{2})$ plane from experimental data, reverse Monte Carlo modeling (RMC), and forward Monte Carlo modeling (FMC). 
    (c) Comparison of the in-plane 2D-$\Delta$PDF extracted from experimental data, reverse Monte Carlo modeling, and forward Monte Carlo modeling.
    }
    \label{fig:rmc}
\end{figure*}

As a brief overview of the crystal structure of ScV$_6$Sn$_6$, Fig. 1(a) shows the undistorted, room-temperature structure.  The bonds between Sn-Sc-Sn along the $c$ axis are highlighted to illustrate the trimer units that collectively contract and displace below $T_{CO}$. Below $T_{CO}$, in the $\sqrt{3}\times\sqrt{3}\times3$ ($R32$) structure, the Sn-Sc-Sn trimers form with displacements that can be differentiated by direction (\textit{up}, \textit{neutral}, or \textit{down}) [Fig. \ref{fig:structure}(b)], with neighboring trimers having dissimilar displacements \cite{tan_abundant_2023}.  The resulting in-plane order found in the $\sqrt{3}\times\sqrt{3}\times3$ structure can be mapped to that of a three-state Potts model with nearest-neighbor antiferromagnetic interactions on a triangular lattice \cite{pokharel_frustrated_2023}. 

The neutral trimer is, however, absent in both the $\sqrt{3}\times\sqrt{3}\times2$ ($P_6/mmm$) and the $2\times2\times2$ ($Immm$) structures predicted to be the most stable by DFT \cite{tan_abundant_2023}. In these cases, the interactions map to a (frustrated) antiferromagnetic triangular lattice Ising model. When interactions are limited to nearest neighbors, the ground state has an extensive degeneracy, and the system is expected to exhibit no long-range order down to the lowest temperatures. Upon including weaker antiferromagnetic second- and third-nearest neighbor terms, however, a wider array of ground-state textures can be realized \cite{smerald_topological_2016}. Notably, the predicted $2\times2\times2$ ($Immm$) structure reflects the ``stripe'' ground state for $J_3 = 0$.
For finite $J_3$ and $J_3/J_2 > 0.5$, the expected ground state is a ``zig-zag'' texture.  This is of particular relevance as the zig-zag structure factor is centered at the $\textbf{q}_H=(\frac{1}{3},\frac{1}{3})$ in-plane wave vector. This is the same position as the short-range charge correlations observed in ScV$_6$Sn$_6$ and suggests $J_1$-$J_2$-$J_3$ as the minimal triangular lattice Ising model necessary to capture its high-temperature charge correlations [Figure 1(c)].

To visualize these correlations, an inverse Fourier transform of the isolated diffuse scattering was performed to generate a $\Delta$PDF map \cite{WeberSimonov+2012+238+247}. This reveals local deviations from the average structure in real space. Fortunately, the observed diffuse scattering is most intense within planes with half-integer $L$ that do not contain any Bragg reflections, and thus the removal of the Bragg scattering (removal of the average structure contributions) is relatively straightforward for this system. The resulting $\Delta$PDF maps from both out-of-plane and in-plane scattering yield clear insights regarding the underlying charge correlations generating this pattern of scattering.


First, we examine the out-of-plane correlations above $T_{CO}$ to test whether they are consistent with the displacive trimer formation present in the low-temperature, charge-ordered structure. Inspection of the scattering along $L$ in the $(H, \frac{4}{3}, L)$ plane [Figure \ref{fig:xyxzPDFcombined}(a)] reveals that the diffuse scattering is absent near $L=0$ and increases in intensity with $L$, consistent with $c$-axis displacements in real space. Figure \ref{fig:xyxzPDFcombined}(b) shows the 3D-$\Delta$PDF extracted from the full scattering volume {after being symmetrized \cite{SuppInfo}}. An inspection of the features along the $z||c$ direction reveals displacement correlations consistent with out-of-phase motion of neighboring trimers along the $c$ axis. Near the $z=\frac{1}{3}c$ and $z=\frac{2}{3}c$ positions, we find dipole-like features typically associated with size-effect correlations. These features are expected to arise from the contraction of the Sc-Sn bond within each trimer, as well as from the fact that the expansion of the Sn1-Sn1 distance associated with neighboring trimers moving away from each other is not equivalent in magnitude to the contraction of the Sn1-Sn1 distance associated with trimers moving towards each other.


\begin{figure*}[htb!]
    \includegraphics[width=1.0\linewidth]{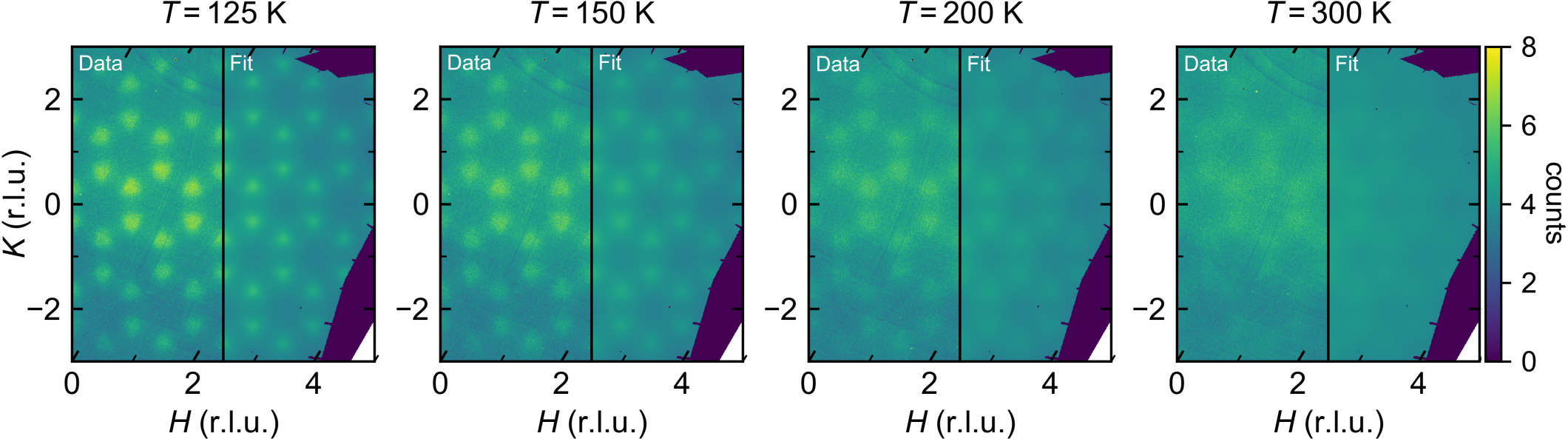}
    \caption{Interaction mean-field model fits to diffuse scattering in the $(H, K, \frac{9}{2})$ plane at various temperatures for $T>T_{CO}$, obtained using \textsc{Spinteract}.}
    \label{fig:spinteract}
\end{figure*}

To verify the interpretation of these correlations, forward Monte Carlo (FMC) modeling was used to calculate the expected scattering and corresponding $\Delta$PDF patterns for trimer formation at high temperature. Only up and down shifted trimers {with correlations along the $c$ axis} were assumed present, displacement amplitudes were assumed equal to those observed in the low-temperature charge-ordered state [Fig. \ref{fig:structure}(b)], and antiferromagnetic (repulsive) interactions between neighboring trimers were enforced along $c$. The simulated $\Delta$PDF shows excellent agreement with the experimental results, replicating features at both integer and $1/3$-type positions [Figure \ref{fig:xyxzPDFcombined}(b)]. Simulations were also performed assuming a rigid Sc-Sn distance, and the feature at $1/3$-type positions became a displacement correlation similar to those observed at integer positions, rather than the dipole-like feature observed in the experimental $\Delta$PDF.  {While this confirms a Sn-Sc-Sn trimer displacement pattern at high temperature, we note here that this qualitative comparison is only weakly sensitive to the overall displacement amplitudes assumed in the FMC model.}

The out-of-plane correlation length can be directly extracted from the decay of the intensity of the $\Delta$PDF with increasing $z$ {(Fig. S1 \cite{SuppInfo})}. Fits were performed to the decay function:  $|\Delta \mathrm{PDF}| = A e^{zc/\xi_c}$
where $A$ is a refined scaling factor, $z$ is the distance along $c$ in lattice units, $c$ is the lattice parameter in $\mathrm{\AA}$, and $\xi_c$ is the correlation length along $c$ in $\mathrm{\AA}$. A correlation length $\xi_c = 13.4(3)~\mathrm{\AA}$ at $T=125~\mathrm{K}$ is obtained \cite{SuppInfo}, in good agreement with the correlation length extracted from the full width at half maximum (FWHM) along $L$ of a Lorentzian fit to the diffuse scattering \cite{pokharel_frustrated_2023}.

Given that the out-of-plane correlation length along the $c$ axis is less than two unit cells, the in-plane charge correlations above $T_{CO}$ were analyzed via a two-dimensional (2D)-$\Delta$PDF transformation of a representative $L=7/2$ scattering plane {without symmetrization \cite{SuppInfo}}.  This plane was chosen because of its lower background intensity and minimal detector artifacts. The extracted 2D-$\Delta$PDF is shown in Fig. \ref{fig:xyxzPDFcombined}(d). The expected antiphase/antiferromagnetic correlations appear at the nearest neighbor positions, but this pattern breaks down beyond two to three unit cells consistent with frustrated interactions. To shed light on the structure of the corresponding real-space correlations, a reverse Monte Carlo (RMC) simulation was performed to refine a supercell initially populated with a random distribution of up and down trimers. The resulting structure after convergence is shown in Fig. \ref{fig:rmc}(a), and resembles a mixture of both zig-zag and stripe patterns at the local scale. 

The calculated scattering for this structure is shown in the right-hand panel of Fig. \ref{fig:rmc}(b), and agrees well with the anisotropic, triangular pattern of scattering from the experimental data. To verify that the observed diffuse scattering is linked to the proposed Ising-like frustration and not simply to a shorter-range version of the low-temperature charge order, reverse Monte Carlo modeling was also performed using all three displacement types (up, down, and neutral) where the frustration is lifted by the addition of a third state {(Fig. S2 \cite{SuppInfo})}. The results fail to converge as closely to the experimentally observed structure factor or to replicate the triangular shape of the diffuse scattering.

Motivated by the observation of the stripe and zig-zag motifs in the Ising-like model, a forward Monte Carlo model was built with antiferromagnetic $J_1$, $J_2$, and $J_3$ [as defined in Fig. \ref{fig:structure}(c)] to search for the corresponding relative energy scales of these interactions. After exploring the parameter space, values near $J_1/k_\mathrm{B}T = -1.25$, $J_2/k_\mathrm{B}T = -0.3$, and $J_3/k_\mathrm{B}T = -0.275$ replicated both the refined real space structure from the RMC simulations [Fig. \ref{fig:rmc}(a)] and the experimental structure factor [Fig. \ref{fig:rmc}(b)]. The weaker diffuse hexagon connecting the diffuse triangles also appears in the experimental data, and it is clearly evident in other diffuse scattering measurements as well \cite{Korshunov2023}. 

To further refine our model, we fit the experimental data to a mean-field model to obtain an absolute energy
scale for the interaction parameters and account for the temperature evolution of the diffuse scattering above $T_{CO}$. Refinements included only Sn1 sites, and accounted for the intra-trimer Sn1-Sn1 interaction $J_{c2}$, the inter-trimer Sn1-Sn1 interaction $J_{c1}$ along $c$, and the in-plane interactions $J_{1}$, $J_{2}$, and $J_{3}$ which were constrained to negative (antiferromagnetic) values. Co-refinements were performed to x-ray data collected at 125~K, 150~K, 200~K, and 300~K. Good overall agreement with the data is obtained, despite the simplicity of the model, as shown in Fig. \ref{fig:spinteract} {and in Fig. S4 \cite{SuppInfo})}. The refined interaction parameters reveal dominant \emph{ferro} intra-trimer coupling $J_{c2}=123$\,K, strongly supporting the hypothesis of parallel Sn1--Sn1 displacements. The inter-trimer coupling $J_{c2}=-52$\,K is antiferro, consistent with the 3D-$\Delta$PDF model shown in Fig.\,\ref{fig:xyxzPDFcombined}(b). The antiferromagnetic in-plane couplings are weaker in magnitude; $J_{1}=-9.3$\,K, $J_{2}=0.0$\,K and $J_{3}=-2.3$\,K, yielding a $J_{3}/J_{1}$ ratio similar to that obtained from forward Monte Carlo simulations. {A forward Monte Carlo model based on these interaction parameters also shows good agreement with the expected in-plane structure factor (Fig. S5 \cite{SuppInfo}).}  The fluctuating, equilibrated model matching the expected thermal evolution contrasts with a number of other inorganic materials with structural degrees of freedom mapped to frustrated spin analogs whose dynamics are frozen \cite{Cairns2016, Fennell2019}. 

The lattice degree of freedom plays the driving role in the formation of charge order \cite{Tuniz2023,Hu2024}, yet the 2D Ising-like structural correlations resolved above $T_{CO}$ seemingly couple to the conduction electrons in a number of ways.   While the Fermi surface is reconstructed primarily in Sn-based bands \cite{Lee2024,PhysRevB.108.245119}, the short-range, frustrated charge order persisting above room temperature along the same in-plane $\textbf{q}=(\frac{1}{3}, \frac{1}{3})$ likely mutes changes in the V-based electronic structure apparent in angle-resolved photoemission measurements through $T_{CO}$ \cite{DiSante2023,Kundu2024}.  The 2D Ising-like charge correlations further correlate to the pseudogap behavior observed in charge transport measurements at temperatures far exceeding $T_{CO}$ \cite{DeStefano2023}, and their higher energy scale is potentially the cause of the anomalous contrast inversion associated with the charge ordered state observed far away from the $E_F$ in scanning tunneling microscopy measurements \cite{Cheng2024}.

{The frustration that arises from a triangular network of lattice displacements mapped to an Ising-like (up/down) degree of freedom in ScV$_6$Sn$_6$ is expected to occur more broadly in a number of other kagome compounds.  Atoms residing either within or nearby the hexagonal voids of kagome nets can generically form chain-like instabilities in the interplane direction whose in-plane correlations are frustrated. The dimerization of Ge atoms reported in FeGe is one example \cite{Miao2023}, and, while this paper was in review, reports showed effects in FeGe similar to those in ScV$_6$Sn$_6$ \cite{korshunov2024pressureinducedquasilongrangesqrt3, subires2024frustratedchargedensitywave}.  For the case of ScV$_6$Sn$_6$, a minimal $J_1$-$J_2$-$J_3$ 2D Ising model on a triangular lattice captures the short-range order and provides an elegant, real-space picture for the fluctuations driving the reported pseudogap behavior.  The surprising agreement with a temperature-dependent mean-field model establishes ScV$_6$Sn$_6$ as a rare platform for studying lattice correlations that map to the 2D Ising model without freezing over a broad range of temperatures above the charge order transition.}

\section{Acknowledgements}

The authors acknowledge fruitful conversations with Matthew Krogstad, William Meier, and Casandra Gomez Alvarado. This work was supported by the U.S. Department of Energy
(DOE), Office of Basic Energy Sciences, Division of Materials Sciences and Engineering
under Grant No. DE-SC0020305.  This work used facilities supported via th the National Science Foundation (NSF) through Enabling Quantum Leap: Convergent Accelerated Discovery Foundries for Quantum Materials Science, Engineering and Information (Q-AMASE-i): Quantum Foundry at UC Santa Barbara (DMR-1906325). Research conducted at the Center for High-Energy X-ray Science (CHEXS) is supported by the NSF (BIO, ENG and MPS Directorates) under Award No. DMR-1829070.
S.J.G.A. acknowledges the additional financial support from the National Science Foundation Graduate Research Fellowship under Grant No. 1650114 and support via the Eddleman Center for Quantum Innovation at UC Santa Barbara.  Use was made of computational facilities purchased with funds from the National Science Foundation (CNS-1725797) and administered by the Center for Scientific Computing (CSC). The CSC is supported by the California NanoSystems Institute and the Materials Research Science and Engineering Center (MRSEC; NSF DMR-2308708) at UC Santa Barbara. Work of J.A.M.P (fits using Spinteract program) was supported by the U.S. Department of Energy, Office of Basic Energy Sciences, Scientific User Facilities Division. Work by B.R.O. was supported by the U.S. Department of Energy (DOE), Office of Science, Basic Energy Sciences (BES), Materials Sciences and Engineering Division.

\bibliography{bibliography}

\end{document}